\documentstyle[epsfig]{mn}
\begin{document}

\title
[On the lack of cold dust in IRAS P09104+4109 and IRAS F15307+3252]
{On the lack of cold dust in IRAS P09104+4109 and IRAS F15307+3252 -- 
their spectral energy distributions and implications for finding dusty AGNs
at high redshift}

\author[J. R. Deane and Neil Trentham] 
{
J. R. Deane$^{1}$ and Neil Trentham$^{2}$\\
$^1$ Physics Department, Queen Mary and Westfield College, Mile End
Road, London E1 4NS \\ 
$^2$ Institute of Astronomy, Madingley Road, Cambridge, CB3 0HA. 
}
\maketitle 

\begin{abstract} 
{ 
We present upper limits on the 850 $\mu$m and 450 $\mu$m fluxes of the
warm hyperluminous (bolometric
luminosity $L_{\rm bol} > 10^{13} \, {\rm L}_{\odot}$)
galaxies IRAS P09104+4109 ($z=0.442$) and IRAS
F15307+3252 ($z=0.926$), derived from measurements using the SCUBA
bolometer array on the James Clerk Maxwell Telescope.  
Hot luminous infrared sources like these are thought to differ from
more normal cold ultraluminous infrared
($L_{\rm bol} > 10^{12} \, {\rm L}_{\odot}$) galaxies in that they derive
most of their bolometric luminosities from dusty AGNs as opposed to
starbursts.
Such hot, dusty AGNs at high redshift are thought to be
responsible for much of the mass accretion of the Universe 
that is in turn responsible for the formation of the 
supermassive black holes seen in the centres of local galaxies. 
The galaxy IRAS P09104+4109 is also unusual in that it is a cD galaxy
in the center of a substantial cooling-flow cluster, not an isolated
interacting galaxy like most ultraluminous infrared 
galaxies. Previously it was known to have large amounts of hot 
($T > 50$ K) dust from $IRAS$ 
observations.  We now show 
that the contribution of cold dust to the bolometric luminosity is
less than 3 per cent.  Most ultraluminous infrared galaxies possess
large amounts of cold dust, and it is now known that some cooling flow
cluster cD galaxies do as well.  Yet this object, which is an extreme 
example of both, does not have enough cold gas to contribute
significantly to the bolometric luminosity.  
We outline physical reasons why this could 
have happened.  We then provide a discussion of stategies for finding 
hot dusty AGNs, given the limitations on submillimetre surveys implied 
by this work. 
}
\end{abstract} 

\begin{keywords}  
infrared: galaxies --
quasars: general -- 
cosmology: observations --
galaxies: individual: IRAS P09104+4109 --
galaxies: individual: IRAS F15307+3252 
\end{keywords} 

\section{Introduction} 

Two extreme ultraluminous infrared galaxies (ULIGs) at redshifts $z<1$
that are thought to derive most of their bolometric luminosity from a
dust-enshrouded active galactic nuclei (AGNs) are IRAS P09104+4109
and IRAS F15307+3252.  These objects differ from optical quasars, even
those with ``infrared bumps'' (e.g.~Sanders et al.~1989), in that they
are very faint at rest-frame optical wavelengths; their $B$-band
luminosities are at least an order of magnitude lower than their
25-$\mu$m luminosities, for example.  These kinds of objects have
attracted much recent attention because of the possibility that much
of the mass accretion in the Universe responsible for producing the
high local density of supermassive black holes (Magorrian et al.~1998)
might have occurred in high redshift analogues of these two galaxies
(Haehnelt, Natarajan \& Rees 1998; Fabian \& Iwasawa 1999; Salucci et
al.~1999).
 
The galaxy IRAS P09104+4109 was identified by Kleinmann et al.~(1988) to
be an ultraluminous IRAS galaxy at $z=0.442$.  It was found
in a follow-up survey of $IRAS$ 60-$\mu$m sources that had non-stellar
colours and no optical counterparts on the Palomar Sky Survey.
Its bolometric
luminosity is $1.4 \times 10^{13} \, {\rm L}_{\odot}$, which is extremely
high, even for a ULIG (it would be classed as a hyperluminous infrared
galaxy by Sanders \& Mirabel 1996).  
Radio, optical, near- and mid-infrared imaging,
spectroscopy, and spectropolarimetry (Kleinmann et al.~1988, Hines \& Wills
1993, Soifer et al.~1996, Green \& Rowan-Robinson 1996; 
Taniguchi et al.~1997, Evans et al.~1998, Hines
et al~1999) 
suggest that this galaxy is
powered by an obscured AGN.  
A hard X-ray detection of IRAS P09104+4109
with {\it BeppoSax} (Franceschini et al.~2000) confirms that this object
contains a powerful AGN. 

The spectral energy distribution (SED) of IRAS P09104+4109 is unusual
in that it peaks between 10 and 30 $\mu$m, implying
that the dust responsible for reradiating the energy from the obscured
AGN must be very hot, in excess of 100 K.  This is very different
from what is seen in other powerful (although lower in luminosity
than IRAS P09104+4109) dust-enshrouded AGNs like the
``warm'' ULIGs Markarian 231 and IRAS 08572+3915 (Sanders et al.~1988a). 
These have dust temperatures closer to 50 K and SEDs which peak longward
of 50 $\mu$m.  The dust in  IRAS P09104+4109 is also considerably hotter
than the dust in many infrared QSOs, like Markarian 1014 (Sanders
et al.~1988b).
That the dust in IRAS P09104+4109 is so much hotter than
the dust in other ULIGs hosting powerful AGNs is perhaps suggestive
that the dust in IRAS P09104+4109 is being heated directly by the AGN 
whereas in the other ULIGs, much of the far-infrared luminosity comes
from colder, more diffuse, dust heated by star formation (Rowan-Robinson
\& Crawford 1989, Rowan-Robinson \& Efstathiou 1993, Rowan-Robinson 2000).

Additionally, IRAS P09104+4109 is a cD galaxy in a rich X-ray cluster
that is undergoing a very substantial cooling flow, about 1000
M$_{\odot}$ yr$^{-1}$ (Fabian \& Crawford 1995, Crawford \& Vanderriest 1996).
In this way it is very different from other ULIGs, which are normally
isolated peculiar galaxies, often showing signs of a recent
interaction (Sanders et al.~1988a,
Clements et al.~1996, Murphy
et al.~1996).  
  
So IRAS P09104+4109 is a very unusual ULIG on three grounds: (i) its
bolometric luminosity is very high; 
(ii) its SED is peaked at short wavelengths,
implying a much higher dust temperature; and (iii) it is not a peculiar
galaxy, but a cD galaxy in the center of
a massive X-ray cluster that is undergoing a cooling flow. 

It is interesting to speculate as to whether these properties are related;
in particular one might ask whether or not the cooling flow can fuel the
AGN and in doing so generate the high bolometric luminosity and dust
temperature.  A related question is: does IRAS P09104+4109 have any cold
($T \leq 50 \, {\rm K}$) dust at all, or is all the dust in this galaxy  
extremely hot and directly associated with the AGN.  Almost all other
ULIGs (Sanders \& Mirabel 1996) and some high-redshift quasars (McMahon
et al.~1999) possess cold gas.  Some cooling flow clusters do too, as
inferred from their SCUBA submillimetre detections (Edge et al.~1999) and CO 
line measurements (e.g.~Bridges \& Irwin 1998).  Cooling flow clusters also
tend to have star-formation in progress at their centers
(Cardiel et al.~1998, Crawford et al.~1999), which may be fuelled 
by this cold gas.  This star formation could could directly produce the dust
seen by Edge and collaborators.
In IRAS P09104+4109, we have the additional evidence
that there exists a high column density of X-ray absorbing gas
(Iwasawa, Fabian \& Ettori 2001) which is likely to be cold.  

The galaxy IRAS F15307+3252 ($z=0.926$; Cutri et al.~1994) is another
hyperluminous infrared galaxy
thought to be powered by a luminous dusty AGN (Hines et al.~1995).
It has somewhat cooler IRAS colours than IRAS P09104+4109 (but still
warmer than the majority of ULIGs), suggesting
its dust is at a lower temperature than that of IRAS P09104+4109. 
This object also differs from IRAS P09104+4109 in that it does not
exist in an X-ray cluster.  In fact, this object has yet to be detected
in X-rays (Fabian et al.~1996, Ogasaka et al.~1997).
It is presumably not associated with a substantial cooling flow.
Unlike IRAS P09104+4109, there is no {\it a priori} reason to expect
cold gas in this galaxy.
 
At high redshifts ($z>2$), objects which contain dusty AGNs are certainly
known, such as IRAS F10214+4724 (Rowan-Robinson et al.~1991) and APM 08279+5255
(Irwin et al.~1998).  However the discovery of both of these objects has been
serindipitous, often helped by strong gravitational lensing (see
Eisenhardt et al.~1996 and Lacy, Rawlings
\& Serjeant 1998 for IRAS F10214+4724 and
Egami et al.~2000 and references therein for APM 08279+5255).
More systematic searches are required if we are to address the issues
highlighted in the opening paragraph of this section.  Surveys done at
X-ray wavelengths offer one approach, and have begun to uncover some
hot dusty objects (Wilman, Fabian \& Gandhi 2000).  But these will not
find X-ray-quiet dusty AGNs like IRAS F15307+3252.  Submillimetre searches
with current or future bolometer arrays offer 
another method, but these will not find dusty AGNs if {\it all} the dust
is hot; unless there is some cold dust present,
sources would then fall below the submillimetre confusion
limit (Blain et al.~1998).  
The presence or absence of cold dust in such objects therefore
has important implications for cosmological surveys. 

We did not detect this cold gas in either object, despite taking very
deep integrations.
In Section 2, we descirbe the observations.
In Section 3, we present the results, compute SEDs, compare
IRAS P09104+4109 and IRAS F15307+3252 to other infrared-luminous AGNs,
discuss the physical processes that may be responsible for generating a
lack of cold gas in these objects, and outline the consequences for
systematic surveys of high-redshift dust-enshrouded AGNs.
In Section 4 we summarize. 
Throughout this work, we assume the following cosmological parameters: 
$H_0 = 65\,\, {\rm km} \, {\rm s}^{-1}
\, {\rm Mpc}^{-1}$,
$\Omega_{\rm matter}=0.3$,
$\Omega_{\Lambda}=0.7$.  

\section{Observations and Data Reduction}

\begin{table*}
\caption{Summary of SCUBA Observations}
{\vskip 0.10mm}
{$$\vbox{
\halign {\hfil #\hfil && \quad \hfil #\hfil \cr
\noalign{\hrule \medskip}
Source & Date & Number of & Total time &
$<\tau_{850\,\mu{\rm m}}>$ & $<\tau_{450\,\mu{\rm m}}>$ &
F$_{850\,\mu{\rm m}}$ & F$_{450\,\mu{\rm m}}$ &\cr
 & & integrations & on source & & &     &     &\cr
 & & & & & & mJy & mJy &\cr
 & & & & & &     &     &\cr
IRAS P09104+4109 & 20 Jan.~1999$^{\mathrm{*}}$ &  50 &  900 s & 0.25 & 1.3 &
      4.24 $\pm$ 3.36  & $-$19.4  $\pm$ 34.2 &\cr
       & 18 Mar~ 2000$^{\mathrm{**}}$ & 120 & 2160 s & 0.26 & 1.4
     & 1.44 $\pm$ 3.61  & $-$13.35 $\pm$ 41.11 &\cr
       &  4 Apr.~2000$^{\mathrm{**}}$ & 280 & 5040 s & 0.39 & 2.2
       & 3.87 $\pm$ 4.18  & $-$29.29 $\pm$ 62.79 &\cr
 & & & & & &     &     &\cr
IRAS F15307+3252 &  6 Feb.~2000$^{\mathrm{**}}$ & 175 & 3150 s & 0.45 & 2.6
      & 3.21 $\pm$ 4.36 &    4.55 $\pm$ 83.24 &\cr
       &  4 Apr.~2000$^{\mathrm{**}}$ &  70 & 1260 s & 0.36 & 2.0
       & 1.79 $\pm$ 4.13 &    6.98 $\pm$ 36.23 &\cr
\noalign{\smallskip }
\noalign{\hrule \smallskip}\cr}}$$}
\begin{list}{}{}
\item[$^{\mathrm{*}}$] Observations performed by K.~Hodapp (University of
Hawaii)
\item[$^{\mathrm{**}}$] Observations in service mode.  
\end{list}
\end{table*}

Both sources were observed with the Submillimetre Common User
Bolometer Array (SCUBA; Holland et al.~1999) on the
James Clerk Maxwell Telescope (JCMT) on Mauna Kea.  
This instrument consists of
two bolometer arrays simultaneously viewing the same region of the sky
through a dichroic beamsplitter, arranged in 
a hexagonal grid with approximately
one telescope-beamwidth separating the detectors.
We used standard filters: the  
``short-wavelength array'' corresponded to an effective wavelength of
observation of $\lambda_0$ = 450 $\mu$m and the
``long-wavelength array'' to $\lambda_0$ = 850 $\mu$m.  

All observations were conducted in photometry mode, in which the
source is positioned onto the central pixel of each array. The data
takes the form of a time-dependent voltage reading from each bolometer
while the telescope secondary mirror chops the source in and out of
the beam, with a throw of 60 arcsec.  In photometry mode, one
``integration'' is defined as a cycle of 9 on-off chopped-pairs
positionally offset by 2 arcsec, for a total of 18 seconds of
observation. Primary and secondary flux calibration sources were
observerd to convert the bolometer voltages into fluxes.  The details
of the observations are summarized in Table 1, where we list
integration times, the local atmospheric optical depth at the time of
observation, and the measured fluxes.

The alternative positive-and-negative waveform of the chopping
secondary was taken out of the voltage stream. 
Next, the time-dependent variations of the sky brightness
were estimated using the signal variations in the bolometers at the
periphery of each array, and then were subtracted from the data
streams.  Two methods of estimating the sky variations (based on using
either the mean or the median of the outer bolometer readings) were
employed, which proved mutually consistent to within 15\% in all
cases.  We then subtracted the mean of these two sky brightness
estimates from the values at the central (on-source) pixel.  In all
cases, this sky subtraction reduced the uncertainty on the mean source
voltage while the improved value remained consistent with the
non-sky-subtracted measurements within the latter's larger
uncertainties.  For each dataset (observing run/source combination),
the mean and the variance of the central pixel voltage over time was
then determined.

This entire procedure was first performed on the observations of the
secondary flux calibrators to determine the flux conversion factor
(FCF), by which the mean voltage (in mV) is converted into the
observed source flux (in mJy).  The following sources were used as
flux calibrators:  Mars, CRL618, IRAS+16293, OH231.8, and IRC+10216.
Their assumed fluxes were taken from the SCUBA calibration web
page
(http:www.jach.hawaii.edu/JACpublic/JCMT/
continuum$_{-}$observing/SCUBA/).
The FCFs were determined independently for each array
(i.e. wavelength) and for each observing run.  Most of the data were
acquired in service mode under ``Category 3'' weather conditions,
which are considered suitable for observations using the Long
Wavelength Array but inadequate for use with the Short Wavelength
Array.  Appropriately, the 850-$\mu$m FCFs were measured to be more
stable than the 450$\mu$m FCFs.  During a single observing run, the
850$\mu$m FCFs typically varied by $\sim$10 $-$ 15\%, and showed
run-to-run variations of $\sim$30\%.  The 450-$\mu$m FCFs varied much
more widely, up to a factor of two or more, both during a single
observing run, and between runs.  The variation of the calibrator FCFs
was one of the most significant contributions to the (systematic)
uncertainty in our results.  
Faint-object photometry at
submillimetre wavelengths is a difficult 
and error-prone exercise, particularly under
below-average weather conditions such as these, and the uncertainty
in our final fluxes (see Table 1) is large.    
Neither IRAS P09104+4109 nor 
IRAS 15307+3252 were detected within our errors: only upper limits
to the source fluxes were obtained.  

In addition to these observations, an upper limit to the 350-$\mu$m
observed-frame (243-$\mu$m rest-frame) flux of IRAS P09104+4109 was
obtained at the Caltech Submillimeter Observatory in October and
December of 1997 using the SHARC camera (Hunter, Benford \& Serabyn
1996).  Although superior weather conditions were experienced, the
collecting area, observing efficiency and stability of the SHARC
camera are less than that of SCUBA.  This is primarily because the
SHARC detector array is one-dimensional and has several inoperative
elements near the array center. 
The upper limit
to the flux obtained ($\sim$300 mJy) does not contribute significantly
to the 450 and 850 limits discussed here.

\section{Results and Discussion}

\begin{figure} 
\begin{center}
\vskip-4mm
\psfig{file=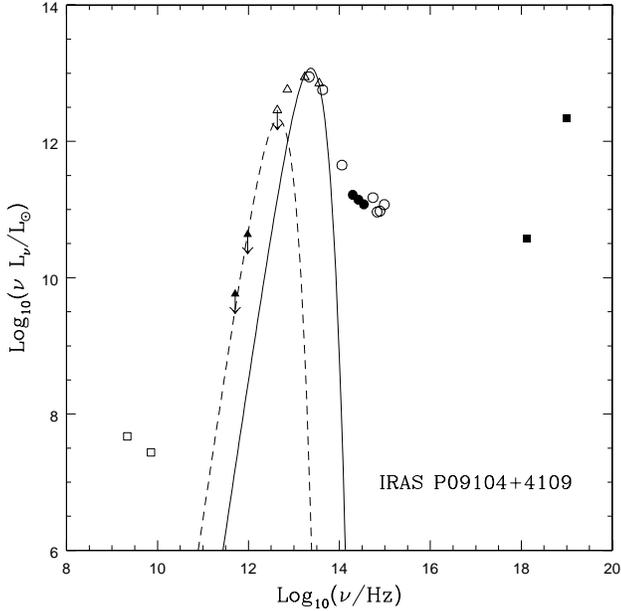, width=8.65cm}
\end{center}
\vskip-3mm
\caption{
The spectral energy distribuion of IRAS P09104+4109.
The points are: open circles 
-- 0.44 $\mu$m, 0.55 $\mu$m, 0.64 $\mu$m, 0.79 $\mu$m,
3.8 $\mu$m, 10.1 $\mu$m and 20 $\mu$m (all in the observer frame) 
from Kleinmann et al.~(1988); open squares 
-- radio continuum at 6 cm and 20 cm from Hines \& Wills (1993);
filled circles -- 1.25 $\mu$m, 1.65 $\mu$m and 2.2 $\mu$m
from Soifer et al.~(1996); 
open triangles --  
12 $\mu$m, 25 $\mu$m, 60 $\mu$m and 100 $\mu$m limit from the  {\it IRAS Faint
Source Catalog};
filled triangles -- 450 $\mu$m and 850 $\mu$m SCUBA 2$\sigma$ limits 
from this work;
filled squares -- 0.5--7 keV {\it Chandra} and 14--40 keV {\it BeppoSax}  
detections from Iwasawa et al.~(2001). 
The solid line is the best-fitting modified (by a dust emissivity 
$k \propto \nu^{1.5}$ law) blackbody fit to the {\it IRAS} detections;
this has a dust temperature $T = 206\,{\rm K}$ and peaks at 13 $\mu$m, but
is a statistically poor fit to the three data points.  
The solid line is the
maximum contribution from a cold (37 K) dust component given the limits
imposed by our SCUBA data. 
}
\end{figure}

\begin{figure}
\begin{center}
\vskip-4mm
\psfig{file=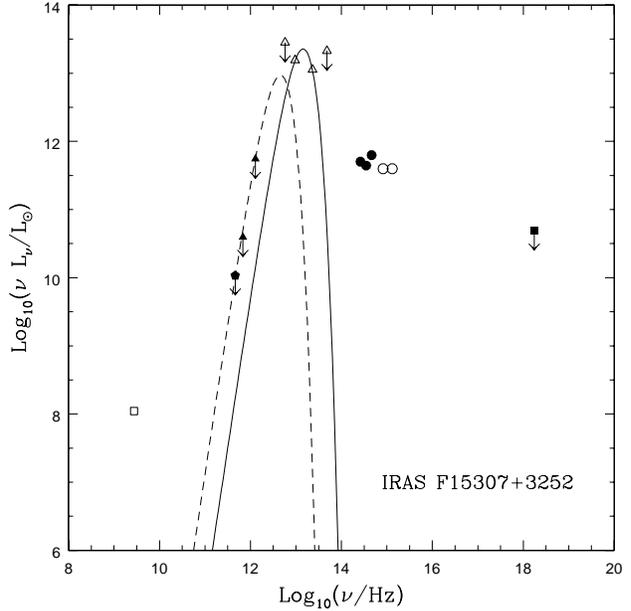, width=8.65cm}
\end{center}
\vskip-3mm
\caption{
The spectral energy distribuion of IRAS F15307+3252 
The points are: open circles
-- 0.44 $\mu$m and 0.70 $\mu$m,
(all in the observer frame)
from Cutri et al.~(1994); open square
-- radio continuum at 21 cm from the FIRST survey of
Becker, White \& Helfand (1995);
filled circles -- 1.25 $\mu$m, 1.65 $\mu$m and 2.2 $\mu$m
from Soifer et al.~(1994);
open triangles --
12 $\mu$m, 25 $\mu$m, 60 $\mu$m and 100 $\mu$m limit from the  {\it IRAS Faint
Source Catalog};
filled triangles -- 450 $\mu$m and 850 $\mu$m SCUBA 2$\sigma$ limits
from this work;
filled pentagon -- 650 $\mu$m lmit from Yun \& Scoville (1998); 
filled squares -- 2 --10 keV {\it ASCA} limit from 
Ogasaka et al.(1997).
The solid line is the best-fitting modified (by a dust emissivity
$k \propto \nu^{1.5}$ law) blackbody fit to the {\it IRAS} detections;
this has a dust temperature $T = 126\,{\rm K}$ and peaks at 21 $\mu$m. 
The solid line is the
maximum contribution from a cold (37 K) dust component given the limits
imposed by the data of Yun \& Scoville (1998).  
}
\end{figure}

Combining data from all the runs listed in Table 1, we find the following
fluxes:
for IRAS P09104+4109, $3.18 \pm 2.12 \, {\rm mJy}$ 
(3$\sigma$ $<$ 9.54 mJy) at 850 $\mu$m and
$-18.77 \pm 24.25 \, {\rm mJy}$ 
(3$\sigma$ $<$ 72.75 mJy) at 450 $\mu$m, and  
for IRAS F15307+3252, $2.46 \pm 3.00 \, {\rm mJy}$ 
(3$\sigma$ $<$ 11.46 mJy) at 850 $\mu$m and   
$6.59 \pm 33.22 \, {\rm mJy}$ 
(3$\sigma$ $<$ 106.3 mJy) at 450 $\mu$m.
For IRAS F15307+3252 the constraints from our results are comparable 
to those from Yun \& Scoville (1998), although at shorter wavelengths. 
SEDs are plotted in Figures 1 and 2. 

\subsection{The lack of cold gas in IRAS P09104+4109}

It is clear from Figure 1 that IRAS P09104+4109 has very little cold dust;
reradiated energy from a cold dust component contributes 
a fraction below 3.0 \% of the bolometric luminosity.  The contribution
to the 850-$\mu$m flux that we would expect from the hot dust responsible
for the 25-$\mu$m $IRAS$ detection accounts for only 1/400 of our observed  
2$\sigma$ upper limit; if this was the {\it only} dust in the galaxy,
it would take several weeks to detect with SCUBA! 

The absence of cold dust in a powerful AGN like this is certainly unusual.
High-redshift optical quasars often have cold dust (McMahon et al.~1999),
so one might expect an infrared quasar like IRAS P09104+4109 to have
some too, particularly since the X-ray measurements (Iwasawa et al.~2001)
tell us that plenty of cold gas is present.

The solution to this apparent paradox is presumably tied up in the way
that gas condenses out of the hot X-ray phase in a cooling flow.
If gas condenses in such a way that there is no new star formation, dust
generation will be inefficient -- carbon atoms associated with this cooling
gas are more likely to interact with four hydrogen atoms and become methane
rather than with numerous other carbon atoms and become dust grains.
This could explain the presence of large amounts of cold X-ray absorbing
gas (Iwasawa et al.~2001)
and no cold dust.  If the cooling flow is also responsible for fuelling the
AGN, any star formation that happens along the way must be confined
to regions very close to the AGN. 
Red giants and supernovae associated with this star formation
could generate the hot dust seen
by $IRAS$.  
 
\subsection{The lack of cold gas in IRAS F15307+3252}

We also did not detect cold dust in IRAS 15307+3252.  In this case
the upper limit on the fraction of the bolometric luminosity that could
come from a cold dust component is 0.3 \%. 

While we had no {\it a priori} reason to expect dust in thie galaxy,
this is still a suprising result.  High-redshift optical quasars often
have cold dust (McMahon et al.~1999).  Yet this galaxy, which is opically
faint (presumably due to internal extinction) and is such a strong
60-$\mu$m emitter, has very little.  This result is in some ways more
perplexing than the non-detection for IRAS P09104+4109 since the $IRAS$
colours are cooler. 

These two non-detections have implications for the discovery of hot,
dusty AGNs at high-redshift using infrared and submillimetre surveys.
We investiate these implications in the rest of this section.

\subsection{Application to high redshift}

In Figure 3, we plot an optical-infrared vs.~infrared-submillimetre 
colour-colour diagram for the sample of dust-enshrouded AGNs presented in
Table 1.  In the rest of this section we will use this diagram to assess
the efficacy of finding dusty AGNs at high redshift by a variety of methods. 
Some particular features worth noting are:
\vskip 1pt \noindent
(i) higher temperature sources are to be found near the top of this
diagram, but this does not tell us immediately that 
all sources in this region
of the diagram are AGN-powered.  For example, IRAS F10214+4724 is to
be found near the top of the diagram yet it may derive a substantial
fraction of its total luminosity from an obscured starburst (Green \&
Rowan-Robinson 1996); 
\vskip 1pt \noindent
(ii) how far to the right of this diagram 
is a measure of how optically bright an object is, which in turn has
implications for our ability to infer the presence of a powerful AGN
using optical spectroscopy.  For example, the ``Q1'', ``Q2'', ``CLOV'',
and ``APM'' points all represent objects with AGNs so powerful that they
likely are responsible for the high bolometric luminosities of the objects; 
\vskip 1pt \noindent
(iii) this diagram gives no indication of the absolute luminosities of
the objects, rather the relative luminosities at different wavelengths.
This has implications for the detectability of the objects.  For example,
APM 08279+5255 is near the top of this diagram yet has been detected
at submillimetre wavelengths with SCUBA (Lewis et al.~1998); this is
becuase both its bolometric
luminosity and gravitational magnification are so high that its
submillimetre flux is still measurable.
Conversely, IRAS P09104+4109, as we saw here, was not detected, although
its spectral energy distribution is perhaps not all that different 
from that of APM 08279+5255 (note
that the increased distance of APM 08279+5255 relative to 
IRAS P09104+4109 is compensated for by the negative K-correction (see
Section 3.3.1);   
\vskip 1pt \noindent
(iv) Radio and X-ray luminosities can be quite different for objects
in similar regions on Fig.~3.  This can also have implications regarding
the detectability of the different objects.  

\subsubsection{Submillimetre surveys}

The main result of the present paper is that the hot dusty AGNs 
IRAS P09104+4109 and IRAS F15307+3252 have very little if any cold gas.
So if {\it all} of the dust-enshrouded accretion in the Universe took
place at high redshift in objects like these, we would not see them
in submillimetre surveys like those carried out with SCUBA (e.g.~Smail,
Ivison \& Blain 1997, Hughes et al.~1998, Cowie \& Barger 2000).  
The only ones we would see in
such surveys are (1) extremely luminous, possibly highly-magnified, 
objects like APM 08279+5255, and (2) the most distant objects at very
high redshifts $z > 6$: for sources with dust emissivity $k_{\nu} \sim
\nu^{1.5}$ (e.g.~Blain et al.~1999), the observed flux $\sim 
(1+z)^{4.5} d_L(z)^{-2}$ where $d_L(z)$ is the luminosity distance, 
so that the negative K-correction
more than compensates for the dimming due to the increased
distance by a significant amount (see Fig.~4).  
However, we do not expect the vast
majority of sources to fall into either of these two categories.  

\subsubsection{Infrared surveys}

Mid- and far-infrared surveys, like those to be carried out with the
$SIRTF$ satellite (http://ipac.caltech.edu/sirtf/) offer a far more
productive strategy for finding hot dusty AGNs at high
redshift: both IRAS P09104+4109 and IRAS F15307+3252 would be
easily detected at redshifts $z<4$.  The absence of submillimetre 
detections at the positions of bright mid-infrared sources would suggest
the presence of hot dusty objects.  Establishing that 
such objects are AGN-powered
will be more difficult, particularly if they are optically too faint
(as both IRAS P09104+4109 and IRAS F15307+3252 would be at high $z$) to
permit spectroscopy with the largest ground-based telescopes.  
However, the analogy with the lower-redshift objects 
studied here in the context of the observations and models outlined
in Section 1 would be suggestive. 

\subsubsection{Optical, radio and X-ray surveys}

Surveys at other wavelengths can also uncover hot dusty AGNs, but in each
case the bias introduced by the selection technique is severe and
complete samples cannot be obtained.  The reason for this is that so
little of the bolometric luminosity of hot, dusty sources emerges outside
the far- and mid-infrared wavebands. 

Optical surveys can uncover extreme objects like APM 08279+5255
(Irwin et al.~1998).  But these represent the rare instances where the
AGN is visible directly at rest-frame ultraviolet wavelenghts, possibly 
because it has burnt through its dust shroud.  There still must be enough dust
present, however, to allow it to be red enough to be identified in these
surveys.  It must also be extremely luminous, and in the case of
APM 08279+5255, highly magnified.  The majority of dusty AGNs at high
redshift may well have properties very substantially different from this
and would consequently be missed in optical surveys.  Both
IRAS P09104+4109 and IRAS F15307+3252 would be missed in optical surveys,
for example; their optical colours and magnitudes are unremarkable. 
  
Radio surveys can also reveal some dusty AGNs like 3C 318 (Willott, Rawlings
\& Jarvis 2000).  But, as for optical AGNs (Goldschmidt et al.~1999), it
seems likely that the majority of such objects are not radio-loud.   

Addtionally X-ray surveys would certainly uncover some hot, dusty
AGNs like IRAS P09104+4109 (although not IRAS F15307+3252).  The
presence of a population of sources seen by {\it Chandra} but not
SCUBA (Fabian et al.~2000; see also the discussion of energy budgets
and the relation to the local MDO density
of Trentham \& Blain 2001 and the more recent observations of
Bautz et al.~2000, Hornschemeier et al.~2000, and
Barger et al.~2001) may have revealed such a population already.
Infrared observations of these sources with $SIRTF$ will tell us whether or
not these objects contain substantial amounts of hot dust.  
If so, the analogy with the local objects
IRAS P09104+4109 and IRAS F15307+3252 outlined in Section 3.2.2 would suggest
that they might be AGN-powered.  Two X-ray sources have already had $ISO$
mid-infrared counterparts identified (Wilman et al.~2000), but longer
wavelengths are required to establish the dust temperatures.
 
\begin{table*}
\caption{Objects plotted in Figure 3} 
{\vskip 0.30mm}
{$$\vbox{
\halign {\hfil #\hfil && \quad \hfil #\hfil \cr
\noalign{\hrule \medskip} 
Notation & Object + Description & Redshift(s)  & Data references  &\cr
 &  (additional to text) & & &\cr
\noalign{\smallskip \smallskip} 
\cr 
P09 & IRAS P09104+4109 & 0.44 &  
      Kleinmann et al.~1988, $IRAS$ Faint Source Catalog, this work &\cr
 & & & &\cr
F15 & IRAS F15307+3252 & 0.93 & 
      Cutri et al.~1994, $IRAS$ FSC, this work, Scoville 1997 &\cr  
 & & & &\cr
F10 & IRAS F10214+4724 & 2.29 & 
      Rowan-Robinson et al.~1993 &\cr   
 & & & &\cr
APM & APM 08279+5255 & 3.87 & 
      Irwin et al.~1998, Lewis et al.~1998, Ibata et al.~1999 &\cr 
 & & & &\cr
CLOV & H1413+117 & 2.56 & 
      Barvainis et al.~1995, Hughes et al.~1997 &\cr 
 & ``Cloverleaf Quasar'' & & &\cr
 & & & &\cr
Q1 & Low-$z$ + high $L$ PG quasars & $0-2$ (mostly $<0.5$) & 
       Sanders et al.~(1989) &\cr 
  & L$_{\rm bol}$ $>$ 10$^{12}$ L$_{\odot}$ & &\cr
 & & & &\cr
Q2 & Low-$z$ + low $L$ PG quasars & $0-2$ (mostly $<0.5$) & 
       Sanders et al.~(1989) &\cr 
  & L$_{\rm bol}$ $<$ 10$^{12}$ L$_{\odot}$ & & &\cr
 & & & &\cr
Q3 & High-$z$ quasars & $3.9-4.7$ &
       McMahon et al.~(1999) &\cr
 & & & &\cr
Q4 & SMM J04135+1027$^{*}$ & 2.83 &
       Knudsen et al.~(2000) &\cr
 & Submillimetre-selected quasar & & &\cr
 & & & &\cr
S1 & SMM J02399$-$0136 & 2.80 & Ivison et al.~1998 &\cr 
S2 & SMM J14011$+$0252 & 2.56 & Ivison et al.~2000 &\cr
S3 & SMM J02399$-$0134 & 1.06 & Soucail et al.~1999, Smail et al.~2000 &\cr
 & & & &\cr
HR10 & ERO J164502+4626.4 & 1.44 & Dey et al.~1999 &\cr
 & (Object 10 in Hu \& Ridgway 1994) & & &\cr
 & & & &\cr
3C & 3C318     & 1.57 & Willott et al.~2000 &\cr
4C & 4C41.17   & 3.80 & Graham et al.~1994, Archibald et al.~2000 &\cr 
8C & 8C1435+63 & 4.25 & Lacy et al.~1994, Archibald et al.~2000 &\cr
 & & & &\cr
CF1 & A2390 cD   & 0.23 & Edge et al.~1999 &\cr
 & cooling flow cluster & & &\cr
CF2 & A1835 cD   & 0.25 & Edge et al.~1999 &\cr
 & cooling flow cluster & & &\cr
 & & & &\cr
WFG & CXOUJ215333.2+174209 & 0.575 & Wilman et al.~2000 (WFG00), L{\'e}monon et
     al.~1998 &\cr 
 & (Object B in WFG00) & & &\cr
 & & & &\cr
S88a & Ultraluminous infrared galaxies & $<0.1$ & Sanders et al.~1988a &\cr
S88b & Warm ultraluminous galaxies & $<0.2$ & Sanders et al.~1988b &\cr 
 & & & &\cr
\noalign{\smallskip } 
\noalign{\hrule \smallskip}\cr}}$$}
\end{table*} 

\begin{figure*}
\begin{minipage}{150mm}
{\vskip-0.95cm}
\begin{center}
\epsfig{file=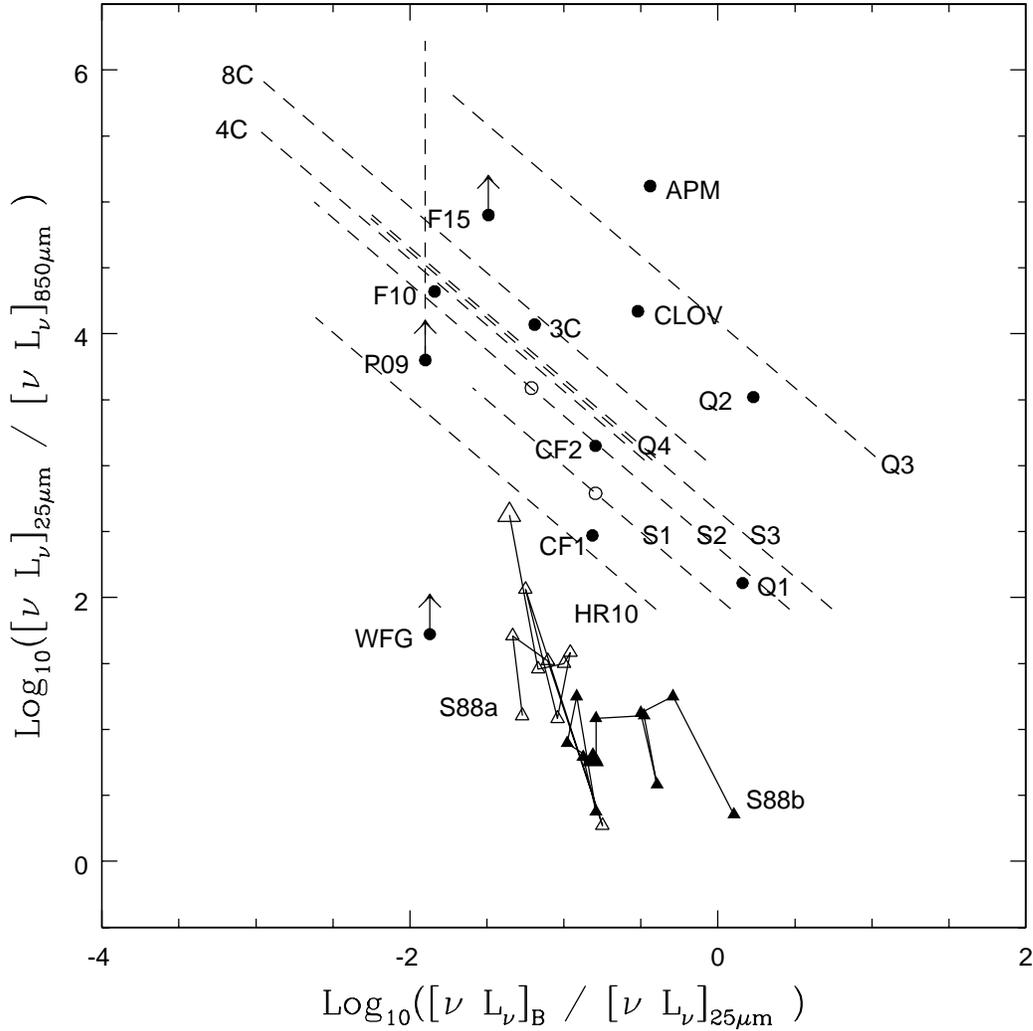, width=14.65cm}
\end{center}
{\vskip-0.35cm}
\caption{Optical ($B$) -- infrared (25 $\mu$m) vs.~infrared (25 $\mu$m) --
submillimetre (850 $\mu$m) colour-colour diagram for the objects listed in
Table 1.
All quantities refer to the rest-frame. The 850-$\mu$m luminosities
are computed assuming a
modified Rayleigh-Jeans spectrum with dust emissivity $k \sim \nu^{1.5}$
(e.g.~Blain et al.~(1999)).
The 25-$\mu$m luminosities
are computed from $IRAS$ measurements, using logarithmic interpolation and/or
extrapolation (but never beyond 130 $\mu$m) if necessary. 
The $B$-band luminosities are computed from
K-corrected optical and near-infrared
fluxes. 
The symbols and labels refer to the objects listed in Table 2.
Additional comments are: 
``P09'' -- The dotted line represents the
permitted values given a lower limit to the 850 $\mu$m flux equal to
the contribution from the hot blackbody that is inferred from the
$IRAS$ 60 $\mu$m and 25 $\mu$m detections -- the solid line in Fig.~1;
``Q3'', ``Q4'', ``S1'', ``S2'', ``S3'', ``HR10'', ``4C'' and ``8C'' -- 
The dotted line represents the locus of permitted values given the 
absence of a mid-infrared rest-frame detection.  
The upper-limit to the 25$-\mu$m
luminosity comes from IRAS measurements, the lower limit from an extrapolation
of the submillimetre SED to mid-infrared wavelengths assuming a modified
Rayleigh-Jeans spectrum as above with dust temperatures of 40 K for
submillimetre galaxies (Blain et al.~1999; Trentham et al.~1999)
and 50 K for quasars (McMahon et al.~1999), 
and the open circle (where present) is derived from   
an $ISO$ 15-$\mu$m detection, converted to a rest-frame  25$-\mu$m flux
assuming a power-law spectrum with index $\alpha = -1.7$ (Blain et al.~2000);
``WFG'' -- the mid-infrared luminosity is 
similarly derived from the $IRAS$ 15-$\mu$m
detection, and the assumed redshift is that derived by
Wilman et al.~(2000) using the HYPERZ population synthesis code
(Bolzonello, Pell{\'o} \& Miralles 2000). 
For ``Q1'', ``Q2'' and ``Q3'', median values
are plotted.  For the ``S88a'' and ``S88b'' sequences, each point (the
open or filled
triangles) represents a galaxy, ordered as in those papers (the large point
represents the first galaxy 
i.e.~Arp 220 for ``S88a'' and IRAS 12071$-$0444 for 
``S88b'') and connected by the lines.   
For IRAS F10214+4724, APM 08279+5255 and the Cloverleaf quasar, the positions
on this diagram have not been corrected for differential magnification (these
are strongly lensed sources) and corrections will shift the points downwards
and to the left if the sizes of the emitting regions increase with
longer wavelength.  For the weakly lensed soures (the three SCUBA galaxies),
differential magnification (Blain 1999) is a small effect.}
\end{minipage}
\end{figure*}

\begin{figure}
\begin{center}
\vskip-4mm
\psfig{file=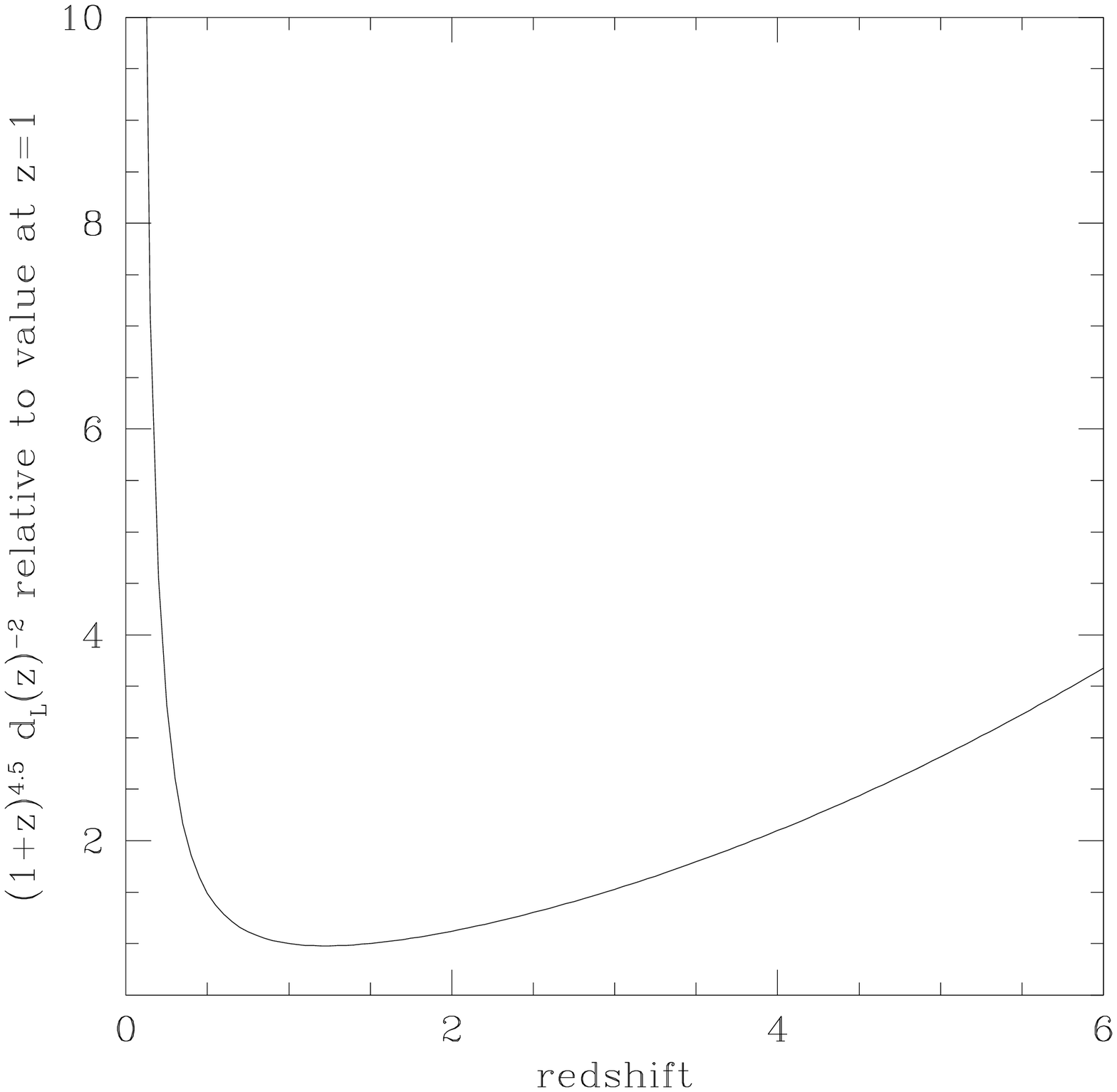, width=8.65cm}
\end{center}
\vskip-3mm
\caption{
The quantity $(1+z)^{4.5} d_L(z)^{-2}$ as a function of redshift $z$.  
}
\end{figure}

\section{Conclusion}

In summary, we have not detected cold,
dusty  gas in either IRAS P09104+4109 or
IRAS F15307+3252.  Dust with a temperature of 40 K, which is what dominates
the bolometric luminosity in most local ULIGs and in the high-redshift
SCUBA sources recently discovered in various surveys (e.g.~Blain  et 
al.~1999) contributes at most 3 \% of the bolometric luminosity of these
objects.

These results were somewhat surprising given that IRAS P09104+4109
is associated with a massive cooling flow in which there is independent
evidence for cold gas (Iwasawa et al. 2001).  
Furthermore, cold dusty gas has been detected in a number of optically
bright quasars (McMahon et al.~1999) so it was unexpected that it would
not be detected in these two objects which are optically faint,
presumably due to internal extinction.  Some
mass of cold dust may still be present,
and may be responsible for some of this internal extinction, but 
it contributes
negligibly to the bolometric luminosity.  
 
These two hot dusty galaxies are thought to derive most of their bolometric
luminosity from AGNs as opposed to starbursts, based on a number of
detailed observational results and models.  Dust-enshrouded AGNs that
are not seen in optical surveys are thought to be responsible for much
of the mass accretion in the Universe that produces the high
local density of supermassive black holes seen in the centres of nearby 
galaxies.  Our results suggest that the best way to find such obejcts
is to look for powerful mid- and far-infrared emitters in $SIRTF$ surveys
that have very low submillimtre SCUBA fluxes.  Some such objects may
have already been seen in joint {\it Chandra}-SCUBA surveys, but there
could also exist a population of hot dusty AGNs 
(like IRAS F15307+3252) that are not X-ray-loud.

\section*{Acknowledgments} 

We would like to thank Klaus Hodapp for obtaining the 1999 data, 
and Nick Tothill for assistance with SURF and calibration issues. 
We also thank Andrew Blain, Aaron Evans, Kazushi Iwasawa and
the Cambridge X-ray group for many helpful discussions.
This research has made use of the NASA/IPAC Extragalactic Database (NED)
which is operated by the Jet Propulsion Laboratory, Caltech, under agreement
with the National Aeronautics and Space Association.

\end{document}